\documentclass[journal]{IEEEtran}
\usepackage{mathtools}
\usepackage{amsmath} 
\usepackage{amssymb}
\usepackage{amsfonts}
\usepackage{verbatim} 
\usepackage{bm,color,soul}
\usepackage{algorithm,algorithmic}

\usepackage{cite}
\usepackage{caption}
\usepackage{stfloats} 
\usepackage{hyperref}
\usepackage{epsfig}
\usepackage{epstopdf}
\usepackage{subfigure}
\usepackage{graphicx}
\usepackage{subeqnarray}
\usepackage{cases}
\usepackage{amsmath}
\usepackage{makecell} 
\usepackage{enumerate}

\newcommand{\rnum}[1]{\uppercase\expandafter{\romannumeral #1\relax}}
\ifCLASSINFOpdf
\else
\fi
\makeatletter
\renewcommand{\maketag@@@}[1]{\hbox{\m@th\normalsize\normalfont#1}}
\makeatother
\usepackage{array}
\usepackage{url}
\usepackage{balance}
\usepackage[T1]{fontenc}

\IEEEoverridecommandlockouts

\begin{document}
\title{Coordinated Power Control for Network Integrated Sensing and Communication}

\author{\IEEEauthorblockN{Yi Huang, Yuan Fang, Xinmin Li, and Jie Xu\vspace{-0.1cm}}
\thanks{Y. Huang is with the Department of Information and Communication Engineering, Tongji University, China (e-mail: huangyi718b@tongji.edu.cn).}
\thanks{Y. Fang and J. Xu are with the School of Science and Engineering (SSE) and the Future Network of Intelligence Institute (FNii), The Chinese University of Hong Kong, Shenzhen, Shenzhen 518172, China (e-mail: fangyuan@cuhk.edu.cn, xujie@cuhk.edu.cn). Y. Fang is the corresponding author.} 
\thanks{Xinmin Li is with the School of Information Engineering, Southwest University of Science and Technology, Mianyang 621000, China (e-mail: lixm@swust.edu.cn). }}

\maketitle \thispagestyle{empty} \vspace{-0.3in}

%
%

\newtheorem{definition}{Definition}
\newtheorem{assumption}{Assumption}
\newtheorem{lemma}{\underline{Lemma}}
\newtheorem{example}{Example}
\newtheorem{theorem}{Theorem}
\newtheorem{proposition}{Proposition}
\newtheorem{conjecture}{Conjecture}
\newtheorem{remark}{Remark}
\newcommand{\mv}[1]{\mbox{\boldmath{$ #1 $}}}

\begin{abstract}
This correspondence paper studies a network integrated sensing and communication (ISAC) system that unifies the interference channel for communication and distributed radar sensing. In this system, a set of distributed ISAC transmitters send individual messages to their respective communication users (CUs), and at the same time cooperate with multiple sensing receivers to estimate the location of one target. We exploit the coordinated power control among ISAC transmitters to minimize their total transmit power while ensuring the minimum signal-to-interference-plus-noise ratio (SINR) constraints at individual CUs and the maximum Cram\'{e}r-Rao lower bound (CRLB) requirement for target location estimation. Although the formulated coordinated power control problem is non-convex and difficult to solve in general, we propose two efficient algorithms to obtain high-quality solutions based on the semi-definite relaxation (SDR) and CRLB approximation, respectively. Numerical results show that the proposed designs achieve substantial performance gains in terms of power reduction, as compared to the benchmark with a heuristic separate communication-sensing design.
\end{abstract}

\begin{IEEEkeywords}
Network integrated sensing and communication (ISAC), coordinated power control, semi-definite relaxation (SDR), Cram\'{e}r-Rao lower bound (CRLB).
\end{IEEEkeywords}

\section{Introduction}
Integrated sensing and communications (ISAC) \cite{OverviewCui, OverviewFan, ZengJLC} has been recognized as one of the candidate techniques for sixth-generation (6G) wireless networks, in which spectrum resources, wireless infrastructures, and communication signals can be reused for the dual role of radar sensing, thus  supporting various new applications such as auto-driving and extended reality. Extensive research efforts have been conducted in the literature to enhance both sensing and communication performances by proposing innovative designs, such as waveform design \cite{Waveform}, transmit beamforming optimization\cite{BeamOpt}, receive signal processing \cite{Receiver}, and resource allocation \cite{OverviewFan}.

While most prior works focused on the ISAC design in the single-cell scenario with one single ISAC transceiver \; (see, e.g., \cite{OverviewFan} and the references therein), recently network ISAC has attracted growing interests, in which multiple ISAC transceivers (e.g., distributed base stations (BSs) in cloud radio access networks (C-RAN)) are enabled to cooperate in performing both distributed radar sensing and coordinated wireless communications (see, e.g., the perceptive mobile network in \cite{PerMobNet}). Network ISAC is expected to bring various advantages over the conventional single-cell  ISAC. From the sensing perspective, network ISAC is able to cover larger surveillance areas, provide better sensing coverage, offer more diverse sensing angles, and capture  richer sensing information  \cite{CramBoundGodrich}. From the communication perspective, different cooperative ISAC transceivers can implement advanced coordinated multi-point transmission/reception (CoMP) techniques to mitigate or even utilize the co-channel interference among different communication users (CUs)\! \cite{CoMP}, and \!also \!properly control the interference between sensing and communication  signals \cite{PerMobNet}. Furthermore, network ISAC provides a viable solution to resolve the full-duplex issue in the single-cell ISAC \cite{Waveform, BeamOpt, Receiver}, by allowing some BSs to act as dedicated sensing receivers.

Despite the benefits, network ISAC imposes new technical challenges in wireless resource allocation, for properly balancing the performance trade-off between sensing versus communication.\; In the literature,\; there have been various prior works investigating the coordinated resource allocation (such as power control and beamforming) for separate sensing\; (e.g., \cite{PowerAlloGodrich}) and communication (e.g., \cite{CoMP}), respectively, but only limited works on that for network ISAC \cite{PerMobNet} \cite{PowerAllocaUAV}. The authors in \cite{PerMobNet} provided an overview on the  perceptive mobile network for network ISAC. \cite{PowerAllocaUAV} studied a multi-unmanned-aerial-vehicle (multi-UAV) network by leveraging UAVs as network ISAC transceivers, in which the UAV location, power allocation, and user association are jointly optimized to maximize the network utility for communication while ensuring the sensing accuracy. 

Different from the prior works, this paper investigates the coordinated power control in a network ISAC system, which consists of multiple ISAC transceivers, sensing receivers, CUs, and one target. In this system, these ISAC transmitters send individual messages to their respective CUs, and at the same time the sensing receivers monitor the target's reflected communication signals for estimating its location. In this way, the network ISAC system integrates the interference channel for communications and the distributed radar sensing into a unified design. Under this setup, we exploit  the coordinated power control at distributed ISAC transmitters to properly balance the performance trade-off between sensing (in terms of Cram\'{e}r-Rao lower bound (CRLB) for location estimation) and communication (in terms of the signal-to-interference-plus-noise ratio (SINR)). In particular, our objective is to minimize  the total transmit power at the ISAC transmitters,  subject to the minimum SINR constraints at individual CUs and the maximum CRLB requirement for target localization. Although the SINR-and-CRLB-constrained power minimization problem is non-convex, we propose two algorithms to obtain efficient solutions, by using the techniques of  semi-definite relaxation (SDR) and CRLB approximation, respectively. Finally, we provide numerical results to validate the performance of our proposed designs, as compared with a benchmark scheme with separate communication-sensing design. It is shown that the SDR-based design outperforms the other two designs, and the CRLB-approximation-based design performs close to the SDR-based design when the SINR requirements become low.

\section{System Model}
\begin{figure}[!h]
\centering
  \includegraphics[width=0.48\textwidth]{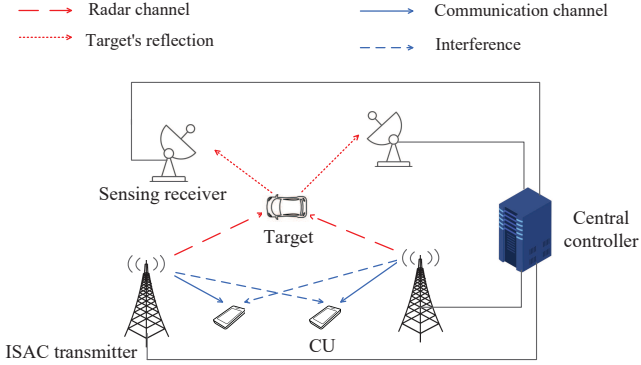}
 \caption{The network ISAC model.}\label{fig:RadarCR}
\end{figure}

We consider a network ISAC system consisting of $M$ ISAC transmitters and $N$ sensing receivers, which are connected to a central controller (e.g., centralized cloud in C-RAN) for joint signal processing. Note that the sensing receivers can either be co-located or separated from the ISAC transmitters. In this system, each ISAC transmitter sends individual messages to one CU, and the transmitted signals are reflected from a target, e.g., a vehicle, and then collected by the sensing receivers for target location estimation. Let $\mathcal{M} = \left\{1, \cdots, M\right\}$ denote the set of ISAC transmitters or CUs, and $\mathcal{N} = \left\{1, \cdots, N\right\}$ denote that of sensing receivers. The coordinates of the $m$-th ISAC transmitter and the $n$-th sensing receiver are denoted as $(\hat{x}_{m}, \hat{y}_{m})$ and $(\check{x}_n, \check{y}_n)$, respectively, $m \in \mathcal{M}$, $n \in \mathcal{N}$. The coordinate of the target is denoted by $(x, y)$. 

First, we consider the resultant interference channel for communication. Let $u_m(t)$ denote the transmit signal by ISAC transmitter $m \in \mathcal M$, which is a random variable with zero mean and unit variance, and $p_m\ge 0$ denote its transmit power. We assume that $u_m(t)$'s are ergodic and independent from each other. Let $h_{m,l}$ denote the channel coefficient from ISAC transmitter $l \in \mathcal M$ to CU receiver $m\in \mathcal M$. Then the received signal at CU $m$ is given by
\begin{align} \label{ReceSig}
y_m(t) = h_{m,m} \sqrt{p_m} u_m(t) + \sum_{l \neq m} {h}_{m,l} \sqrt{p_l} u_l(t) + z_m(t),
\end{align}
where $z_m(t)$ denotes the noise at the receiver of CU $m$, that is a circularly symmetric complex Gaussian (CSCG) random variable with zero mean and variance $\sigma_m^2$, i.e., $z_m(t) \sim \mathcal{CN}(0,\sigma_m^2)$. The corresponding SINR is expressed as
\begin{align}  \label{SINR}
{\rm \gamma}_{m}\left(\boldsymbol{p}\right)= \frac{\left| h_{m,m}\right| ^2 p_m}{ \sum\limits_{l \neq m} \lvert h_{m,l} \rvert^2 p_l  + \sigma_m^2},  \forall  m \in  \mathcal{M},
\end{align}
where $\boldsymbol{p} = [p_1,\cdots, p_M]^T$ with $(\cdot)^{T}$ representing the transpose operator.

Next, we consider the distributed radar sensing, in which  the communication signals $u_{m}(t)$'s are reused for  sensing. Suppose that the radar processing is implemented over an interval $\mathcal{T}$ with duration $T$, which is sufficiently long so that $\int_{\mathcal{T}}\lvert u_{m}(t)\rvert^2dt = T \cdot \mathbb{E}\left[\lvert u_{m}(t)\rvert^2\right] = T$ and $ \int_{\mathcal{T}} u_{m}(t)u^*_{l}(t)dt=T \cdot  \mathbb{E}\left[u_{m}(t)u^*_{l}(t)\right] = 0$, $\forall m \neq l$. It is assumed that different ISAC transmitters and sensing receivers are synchronized in both time and frequency, as commonly assumed in distributed radar \cite{CramBoundGodrich}, which can be implemented in cellular networks via clock calibration through backhaul links or synchronization signals. Then the received signal by the $n$-th sensing receiver is expressed as
\begin{align} \label{ReflectedRadar}
r_{n}(t) = \sum_{m=1}^M h_{n,m}\sqrt{p_m}u_m(t-\tau_{n,m}) + w_{n,m}(t),
\end{align}
where $w_{n,m}(t)$ denotes the noise at sensing receiver $n$,  that is a CSCG random sequence with zero mean and autocorrelation function $\sigma^2_w\delta(\tau)$. $h_{n,m}$ is a coefficient capturing the effects of the target radar cross section (RCS) and the pathloss for the radar propagation path between the $m$-th ISAC transmitter and the $n$-th sensing receiver. Here, we assume that $h_{n_1,m_1}$ is independent from  $h_{n_2,m_2}$ when $m_1 \neq m_2$ and $n_1 \neq n_2$. $\tau_{n,m}$ denotes the propagation delay for radar channel $(n,m)$, i.e., 
\begin{align} \label{Delay}
\tau_{n,m} &= \frac{\sqrt{(\hat{x}_m\!-\!x)^2\!+(\hat{y}_m\!-\!y)^2} +\!\! \sqrt{(\check{x}_n\!-\!x)^2+(\check{y}_n\!-\!y)^2}}{c}  \nonumber\\
&\triangleq \frac{R_{tx}+R_{rx}}{c},
\end{align}
where $c$ is the speed of electromagnetic wave.

For distributed radar sensing, the target's location $(x,y)$ and the channel parameters $\zeta_{m,n}$'s are unknowns to be estimated. It has been shown in \cite{CramBoundGodrich, PowerAlloGodrich} that the CRLB matrix on $x$ and $y$ is given by 
\begin{align}
	\boldsymbol{C}_{x,y}(\boldsymbol{p})= \left\{ \sum_{m=1}^{M-1} p_{m} \left[\begin{matrix}  g_{a_m} & g_{c_m} \\ g_{c_m} & g_{b_m}  \end{matrix}\right]  \right\}^{-1},
\end{align}
where $g_{a_m}$, $g_{b_m}$, and $g_{c_m}$ are respectively defined by 
\begin{align} \label{Parameter}
	g_{a_m} &= \xi_{m} \sum_{n=1}^{N} \lvert h_{n,m}\rvert^2 \left( \frac{\hat{x}_m-x}{R_{tx}}+ \frac{\check{x}_n-x}{R_{rx}} \right)^2, \\
	g_{b_m} &= \xi_{m} \sum_{n=1}^{N} \lvert h_{n,m}\rvert^2 \left( \frac{\hat{y}_m-y}{R_{tx}}+ \frac{\check{y}_n-y}{R_{rx}} \right)^2, \\
	g_{c_m} &= \xi_{m} \sum_{n=1}^{N} \lvert h_{n,m}\rvert^2 \left( \frac{\hat{x}_m-x}{R_{tx}}+ \frac{\check{x}_n-x}{R_{rx}} \right) \nonumber\\
	&\quad\cdot\left( \frac{\hat{y}_m-y}{R_{tx}}+ \frac{\check{y}_n-y}{R_{rx}} \right), m \in \mathcal{M},
\end{align}
with $\xi_{m} = \frac{8\pi^2\beta_m^2 T}{\sigma_w^2c^2}$. Here, $\beta_{m}$ denotes the effective bandwidth of $u_m(t)$, satisfying $\beta_{m}^2= \frac{\int_{B_{m}}f^2\lvert U_m(f)\rvert^2 df}{\int_{B_{m}}\lvert U_m(f)\rvert^2 df}$ with $B_{m}$ denoting the bandwidth of $u_ {m}(t)$ and $U_m(f)$ denoting the frequency domain transformation of $u_m(t)$. Accordingly, the sum of CRLBs for estimating $x$ and $y$ is expressed as 
\begin{align} \label{Cramer-Rao bound}
\sigma^2_{x,y} (\boldsymbol{p}) = {\rm tr}\left(\boldsymbol{C}_{x,y}(\boldsymbol{p})\right) = \frac{\boldsymbol{b}^T\boldsymbol{p}}{\boldsymbol{p}^T\boldsymbol{A}\boldsymbol{p}},
\end{align}
where $\boldsymbol{b} = \boldsymbol{g}_a + \boldsymbol{g}_b$, $\boldsymbol{A} = \boldsymbol{g}_a\boldsymbol{g}_b^T-\boldsymbol{g}_c\boldsymbol{g}_c^T$, $\boldsymbol{g}_a=[g_{a_1}, g_{a_2}, \cdots, g_{a_{M}}]^T$, $\boldsymbol{g}_b=[g_{b_1}, g_{b_2}, \cdots, g_{b_{M}}]^T$, and $\boldsymbol{g}_c=[g_{c_1}, g_{c_2}, \cdots, g_{c_{M}}]^T$. It is assumed  that the target location is roughly  known {\it a priori}, and thus we can optimize the corresponding CRLB to enhance the accuracy for real-time estimation, similarly as in \cite{CramerRaoOpt}.

In particular, we are interested in minimizing the total transmit power of the ISAC transmitters, while ensuring the minimum SINR requirement $\Gamma_m$ at each CU $m\in\mathcal M$ for communication, and the maximum CRLB constraint $\tau$ for target location estimation. The SINR-and-CRLB-constrained power minimization problem is formulated as 
\begin{subequations}
  \begin{align}\label{Opt1}
    &(\text{P1}):&\min_{\boldsymbol{p}}\quad&  \boldsymbol{1}^T \boldsymbol{p} \\
     & & \text{s.t.} \quad& p_m \ge 0,  \forall m \in \mathcal{M},  \\
     & & & \frac{\left| h_{m,m}\right| ^2 p_m}{ \sum\limits_{l \neq m} \lvert h_{m,l} \rvert^2 p_l  + \sigma_m^2} \ge \Gamma_m, \forall m \in \mathcal{M},  \label{Opt1_cons1} \\
     & & & \frac{\boldsymbol{b}^T\boldsymbol{p}}{\boldsymbol{p}^T\boldsymbol{A}\boldsymbol{p}} \le \tau,\label{Opt1_cons2}
  \end{align}
  \end{subequations}
where $\boldsymbol{1}$ is an all-one vector with proper dimensions. For notational convenience, problem (P1) can be equivalently reformulated as the following non-convex quadratic problem. 
\begin{subequations}
  \begin{align}\label{Opt2}
	    &(\text{P2}):&\min_{\boldsymbol{p}}\quad&  \boldsymbol{1}^T \boldsymbol{p} \\
     & & \text{s.t.} \quad&  p_m \ge 0,  \forall m \in \mathcal{M}, \label{opt2_cons1}  \\
     & & & \boldsymbol{g}_m^T  \boldsymbol{p} \geq \Gamma_m \sigma_m^2, \forall m \in \mathcal{M}, \label{opt2_cons2}  \\
     & & &  \boldsymbol{b}^T\boldsymbol{p} - \tau{\boldsymbol{p}^T\boldsymbol{A}\boldsymbol{p}} \le 0, \label{opt2_cons3}
  \end{align}
  \end{subequations}
where $\boldsymbol{g}_m \!=\! [-\Gamma_m\lvert h_{m,1}\rvert^2, \cdots, \lvert h_{m,m}\rvert^2, \cdots, -\Gamma_m\lvert h_{m,M}\rvert^2]^T$. Notice that problem (P1) or (P2)  is a non-convex optimization problem, since the constraint in (10d) (or (11d)) is non-convex.

Before proceeding to solve the problem, we first check its feasibility. Notice that if there is a positive power vector $\overline{\boldsymbol{p}}$ satisfying the SINR constraints in \eqref{opt2_cons2}, then  we can always find a scaling factor $\eta \ge {\rm max} (\frac{\boldsymbol{b}^T \overline{\boldsymbol{p}}}{ \tau \overline{\boldsymbol{p}}^T\boldsymbol{A} \overline{\boldsymbol{p}} }, 1)$ and accordingly set the transmit powers as  $ \tilde{\boldsymbol{p}} = \eta \overline{\boldsymbol{p}}$, which can satisfy the SINR constraints in  \eqref{opt2_cons2} and the CRLB constraint in \eqref{opt2_cons3} at the same time. Therefore, problem (P2) is feasible if and only if there exists a transmit power vector satisfying the SINR constraints in (11c). Then we can check the feasibility of (P2) or (P1) by solving the following linear program via standard convex solvers such as CVX \cite{CVX}. 
\begin{subequations}
  \begin{align}\label{Opt3_1}
	    &(\text{P3}):&\text{Find} \quad& \boldsymbol{p} \\
     & & \text{s.t.} \quad&  p_m \ge 0,  \forall m \in \mathcal{M},  \\
     & & & \boldsymbol{g}_m^T  \boldsymbol{p} \geq \Gamma_m \sigma_m^2, \forall m \in \mathcal{M}. \label{opt3_1_cons2}
  \end{align}
  \end{subequations}
In the sequel, we focus on the case when problem (P1) or (P2) is feasible.

\section{Proposed Solution to Problem (P1)} 

In this section, we propose two algorithms to solve problem (P1) by using the techniques of SDR and CRLB approximation, respectively.

\subsection{SDR-Based Solution}
Based on the SDR technique \cite{SDR,SDRboxCons}, we first define $\boldsymbol{P} \triangleq \boldsymbol{p}\boldsymbol{p}^T$ and
$$
\boldsymbol{Y} \triangleq \left[1 \ \ \boldsymbol{p}^T\right]^T \left[1 \ \ \boldsymbol{p}^T\right] = \left[\begin{array}{cc}
	1 & \boldsymbol{p}^{T} \\
	\boldsymbol{p} & \boldsymbol{P}
\end{array}\right], 
$$
where 
\begin{align}
\boldsymbol{Y} \succcurlyeq 0, \ {\rm rank}(\boldsymbol{Y})=1 . \label{RankCons}
\end{align}
Then, minimizing the objective function $\boldsymbol{1}^T \boldsymbol{p}$ in (\text{P2}) is equivalent to minimizing  $\left|\boldsymbol{1}^T \boldsymbol{p}\right|^2=\left[0,\boldsymbol{1}^T\right]\boldsymbol{Y}\left[0,\boldsymbol{1}^T\right]^T$.  Besides, the left-hand-side of the CRLB constraint \eqref{opt2_cons3} is rewritten as
\begin{align}
\boldsymbol{b}^T\boldsymbol{p} - \tau{\boldsymbol{p}^T\boldsymbol{A}\boldsymbol{p}}&=
{\rm tr}\left({\boldsymbol{b}^T\boldsymbol{p} - \tau{\boldsymbol{p}^T\boldsymbol{A}\boldsymbol{p}}}\right)\nonumber\\
&={\rm tr}\left({\boldsymbol{b}^T\boldsymbol{p} - \tau{\boldsymbol{A}\boldsymbol{P}}}\right)\nonumber\\
&={\rm tr}\left(\left[\begin{matrix}0 &\frac{1}{2} \boldsymbol{b}^T\\ \frac{1}{2} \boldsymbol{b}& -\tau \boldsymbol{A} \end{matrix}\right]\boldsymbol{Y}\right),
\end{align}
where $\mathrm{tr}(\cdot)$ denotes the trace operator. Therefore, constraint \eqref{opt2_cons3} is equivalent to the following constraint:
\begin{align}
{\rm tr}\left(\left[\begin{matrix}0 &\frac{1}{2} \boldsymbol{b}^T\\ \frac{1}{2} \boldsymbol{b}& -\tau  \boldsymbol{A} \end{matrix}\right]\boldsymbol{Y}\right) \leq 0. \label{CRBLrelax}
\end{align}
Furthermore, the SINR constraint \eqref{opt2_cons2} is equivalent to
\begin{align}
[-\boldsymbol{\tilde{\gamma}},  \boldsymbol{G}] \boldsymbol{Y} \boldsymbol{e}_i \geq \boldsymbol{0}, i=1,\cdots, M+1,  \label{SINRrelax}
\end{align}
where $\boldsymbol{e}_i$ is an $(M+1)\times 1$ vector with its $i$-th element being one and others zero, $\boldsymbol{G}  = [\boldsymbol{g}_1, \boldsymbol{g}_2, \cdots, \boldsymbol{g}_M]^T$, and
$\boldsymbol{\tilde{\gamma}} = [\Gamma_1 \sigma_1^2, \Gamma_2 \sigma_2^2, \cdots, \Gamma_m \sigma_M^2]^T$. Accordingly, problem (\text{P2}) is equivalently transformed as

\begin{subequations}
  \begin{align}
    &(\text{P4}): &\min_{\boldsymbol{Y}}\quad&  \left[0,\boldsymbol{1}^T\right]\boldsymbol{Y}\left[0,\boldsymbol{1}^T\right]^T    \\
     & & \text{s.t.} \quad& \boldsymbol{Y} \boldsymbol{e}_i \ge \boldsymbol{0},   i = 1, \ldots, M+1, \\
     & & &\boldsymbol{Y}(1,1) = 1, \\
     & & & \eqref{RankCons},  \eqref{CRBLrelax}, \eqref{SINRrelax}. \nonumber 
  \end{align}
  \end{subequations}
Notice that (P4) is still not convex due to the rank-one constraint in \eqref{RankCons}. To tackle this issue, we use the SDR technique to remove the rank-one constraint, and accordingly get the SDR version of (P4), denoted by (SDR4).  Note that problem (\text{SDR4}) is a convex problem that can be solved by convex optimization tools such as CVX \cite{CVX}. Let $\boldsymbol{Y}^{\star}$ denote the obtained optimal solution to (SDR4), which, however, is of high rank in general. 

Next, we construct a rank-one solution to problem (P2) by Gaussian randomization \cite{SDR} based on the obtained $\boldsymbol{Y}^{\star}$. Let $\boldsymbol{P} = \boldsymbol{Y}_{[2:M+1,2:M+1]}^{\star}$ represent the extraction of rows from $2$ to $M+1$ and columns from $2$ to $M+1$ from the matrix $\boldsymbol{Y}^{\star}$,  for which the eigenvalue decomposition (EVD) is expressed as $\boldsymbol{P} = \boldsymbol{V} \boldsymbol{D} \boldsymbol{V}^T$, where $\boldsymbol{V}\boldsymbol{V}^T =\boldsymbol{V}^T \boldsymbol{V} =\boldsymbol{I}$ and $\boldsymbol{D}$ is a diagonal matrix with non-negative diagonal elements. Then we generate a random vector $\boldsymbol{z}$ as $\boldsymbol{z} = {\rm abs}\left(\boldsymbol{V}\sqrt{\boldsymbol{D}}\boldsymbol{w}\right)$, where $\boldsymbol{w}$ is a Gaussian random vector with mean $\boldsymbol{0}$ and covariance $\boldsymbol{I}$, i.e., $\boldsymbol{w}\sim \mathcal{N}(\boldsymbol{0},\boldsymbol{I})$, and ${\rm abs}\left(\boldsymbol{\cdot}\right)$ denotes the absolute value operator. However, the obtained $\boldsymbol{z}$ may not be feasible for problem (\text{P2}). To deal with this issue, we introduce a scaling factor $\xi >0$ and find a feasible transmit power vector  $\xi \boldsymbol{z}$ for problem (P2) as follows. By substituting $\boldsymbol{p}$ as $\xi \boldsymbol{z}$ in (P2), we find a desirable $\xi$ by solving the following optimization problem:
\begin{subequations}
  \begin{align}
    &(\text{P5}): &\min_{\xi}\quad& \xi\boldsymbol{1}^T \boldsymbol{z} \\
     & &  \text{s.t.} \quad& \xi > 0 \label{opt5_cons1}, \\
     & & &\boldsymbol{g}_m^T \cdot \xi \boldsymbol{z} \geq \Gamma_m \sigma_m^2, \forall m \label{opt5_cons2} \in \mathcal{M},\\
     & & &\tau \xi \ge \frac{\boldsymbol{b}^T\boldsymbol{z}}{ 	\boldsymbol{z}^T\boldsymbol{A}\boldsymbol{z}} \label{opt5_cons3}.
  \end{align}
 \end{subequations}
 Notice that $\boldsymbol{1}^T \boldsymbol{z} $ is always positive, and as a result, the optimal solution $\xi$ to problem (P5) can be obtained as the minimum one that satisfies the constraints in (18b-18d). Therefore, we have the optimal solution to problem (P5) as $\xi^{*} =  \max(\frac{{\Gamma}_1\sigma_1^2}{\boldsymbol{g}_1^{T}\boldsymbol{z}},\cdots,\frac{\Gamma_M\sigma_M^2}{\boldsymbol{g}_M^{T}\boldsymbol{z}},\frac{\boldsymbol{b}^T\boldsymbol{z}}{\tau	\boldsymbol{z}^T\boldsymbol{A}\boldsymbol{z}})$ and obtain a feasible transmit power solution as $\xi^* \boldsymbol{z}$. 
 
It is worth noting that  problem (P5) may not always be feasible. As a result, we need to implement the Gaussian randomization multiple times in general, and accordingly choose the power vector that achieves the minimum total transmit power as the final solution. 

\subsection{CRLB-Approximation-Based Solution}
This subsection presents an alterative solution to problem (P2), motivated by the CRLB approximation in \cite{PowerAlloGodrich}, where the non-convex CRLB constraint \eqref{opt2_cons3} is relaxed as $\boldsymbol{b} - \tau\boldsymbol{A}\boldsymbol{p} \le 0$. Accordingly, problem (P2) is relaxed as the following convex problem that is optimally solvable via CVX.
\begin{subequations}
  \begin{align}
    &(\text{P6}): &\min_{\boldsymbol{p}}\quad&  \boldsymbol{1}^T \boldsymbol{p} \\
     & & \text{s.t.} \quad& p_m \ge 0,  \forall m \in \mathcal{M}, \\
     & & &\boldsymbol{g}_m^T \cdot \boldsymbol{p} \geq \Gamma_m \sigma_m^2, \forall m  \in \mathcal{M},\\
     & & &\boldsymbol{b} - \tau\boldsymbol{A}\boldsymbol{p} \le 0 \label{opt6_cons3}.
  \end{align}
 \end{subequations}
 Let the obtained optimal solution to problem (P6) be  denoted as $\boldsymbol{p}^\star$. Then we use an iterative algorithm to find an efficient solution to the original problem (\text{P2}), in which $\boldsymbol{p}^\star$ is adopted as the starting point $\boldsymbol{p}^{(0)} = \boldsymbol{p}^\star$, and  $\Delta p$ is defined as a step size. In each iteration $i \ge 1$, we first generate $M$ new candidate power vectors $\boldsymbol{z}^{(i-1)}_m$'s, $\forall m\in \mathcal M$, where $\boldsymbol{z}^{(i-1)}_m$ is obtained by subtracting $\Delta p$ from the $m$-th element of $\boldsymbol{p}^{(i-1)}$. Next, we check the feasibility of each candidate power vector (i.e., whether it satisfies the SINR and CRLB constraints), and compare their correspondingly achieved CRLB (as they achieve the same total power) to find the one that minimizes the CRLB, which is then updated to be $\boldsymbol{p}^{(i)}$ for maximizing the sensing performance. The above iterations will terminate until the resultant CRLB approaches the threshold $\tau$.

\section{Numerical Results}
This section provides numerical results to validate the effectiveness of our proposed algorithms. In the simulation, we set the carrier frequency as $6$ GHz and the bandwidth as $1$ MHz. We adopt the Rician channel model with the K-factor being $5$ dB. We also set the noise power spectrum density as $-174$ dBm/Hz, and the SINR constraints to be $\Gamma_m = \Gamma$.

For performance comparison, we consider a benchmark scheme with separate communication-sensing design. In this benchmark, we first optimize the coordinated power control $\boldsymbol{p}$ to minimize the total power $\boldsymbol{1}^T \boldsymbol{p}$, while ensuring the communication-related constraints in \eqref{opt2_cons1} and \eqref{opt2_cons2}, for which the optimal solution is denoted as $\hat{\boldsymbol{p}}$. Next, we scale $\hat{\boldsymbol{p}}$ by a factor $\eta$ to meet the sensing CRLB requirements. The optimal scaling factor is $\eta^{*} = \max(1,\frac{\boldsymbol{b}^T\hat{\boldsymbol{p} }}{\tau	\hat{\boldsymbol{p} }^T\boldsymbol{A}\hat{\boldsymbol{p} }})$ to minimize the sum power. Accordingly, the power control vector is obtained as $\eta^*\hat{\boldsymbol{p} }$.


First, we consider the case when there are two ISAC transmitters, namely ISAC transmitters 1 and 2, which are located at $[-50,0]$ and $[0,50]$ meters~(m), respectively. The coordinates of CU receivers 1 and 2 are $[-20,0]$m and $[20,0]$m, respectively. There are two sensing receivers located at $[-50,-10]$m and $[50,10]$m, respectively. The location of the target is $[30,0]$m  unless stated otherwise.
\begin{figure}[h]
\centering
\vspace{-0.4cm}
  \includegraphics[width= 0.43\textwidth]{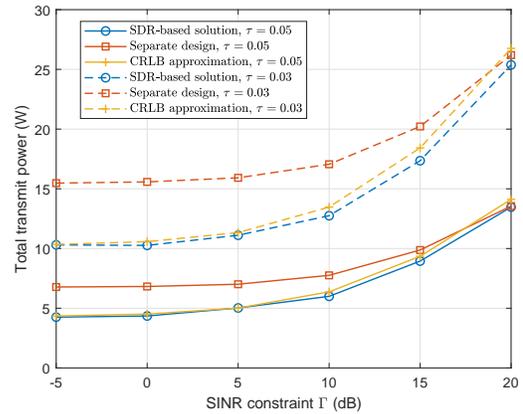}
  \setlength{\belowcaptionskip}{-5pt}
 \caption{The total transmit  power versus the SINR constraint $\Gamma$  in the case with two ISAC transmitters.}
 \DeclareGraphicsExtensions.
 \label{SumPowerVsSINRallUsers_SdrHeuCr_2BS}
\end{figure}

\begin{figure}[h]
\centering
\vspace{-0.5cm}
  \includegraphics[width= 0.43\textwidth]{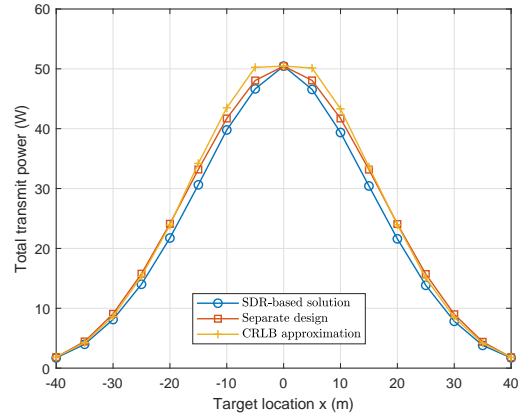}
  \setlength{\belowcaptionskip}{-5pt}
 \caption{The total transmit  power versus the location of target in the case with two ISAC transmitters, where $\tau = 0.05$ and $\Gamma = 10 \rm{dB}$.}
 \DeclareGraphicsExtensions.
 \label{SumPowerVsLocation_2BS}
\end{figure}

Fig.~\ref{SumPowerVsSINRallUsers_SdrHeuCr_2BS} shows the total transmit power versus the SINR constraint $\Gamma$  by considering two different CRLB requirements, where the CRLB thresholds are set as $\tau=0.03$ and $0.05$, respectively. It is observed that the SDR-based solution outperforms both the CRLB-approximation-based solution and the separate design, and the performance gain becomes more significant when the SINR requirement becomes high. It is also observed that the sum transmit power almost remains unchanged when SINR requirement is low (e.g., from $-5$dB to $0$dB). This is because in this case the SINR requirement is easy to be ensured, and thus the total power consumption mainly depends on the given CRLB requirement. Furthermore, the separate design benchmark is observed to achieve the worst performance when the SINR requirement is low, but perform close to the SDR-based solution when it becomes high. This is due to the fact that in the latter case, the SINR constraints become dominant and the CRLB constraints may become inactive, and as a result, the separate design becomes similar to the proposed SDR-based solution. 

Fig.~\ref{SumPowerVsLocation_2BS} shows the total transmit power of ISAC transmitters versus the  horizontal coordinate $x$ of the target, where its location is set as $[x,0]$m. It is observed that the total  transmit power decreases as the target gets closer to one of the ISAC transmitters. It is also observed that the highest transmit power appears when the target is located  near the middle of two ISAC transmitters.

\begin{figure}[htp]
  \centering
  \vspace{-0.4cm}
    \includegraphics[width= 0.43\textwidth]
    {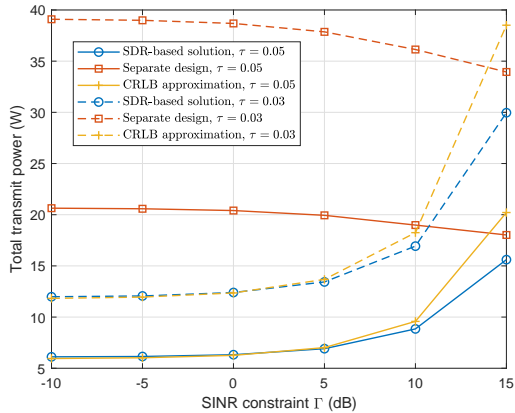}
    \setlength{\belowcaptionskip}{-5pt}
   \caption{The total transmit  power versus the SINR constraint $\Gamma$  in the case with three ISAC transmitters.}
   \DeclareGraphicsExtensions.
   \label{SumPowerVsSINRallUsers_SdrHeuCr_3BS}
  \end{figure}
  
Next, we consider the other scenario with three ISAC transmitters and two sensing receivers. The locations  of three ISAC transmitters are set as $[-100,0]$m, $[100,0]$m, and $[0,100]$m, respectively. The coordinates of CU receivers 1, 2, and 3 are $[-80,20]$m, $[80,20]$m, and $[0,80]$m, respectively. There are two sensing receivers located at $[-100,50]$m and $[100,50]$m, respectively. The location of the target is $[0,50]$m. 

Fig.~\ref{SumPowerVsSINRallUsers_SdrHeuCr_3BS} shows the total  transmit power versus SINR constraint $\Gamma$, where we consider two  CRLB thresholds with $\tau=0.03$ and $\tau=0.05$, respectively. Similar observations are made in Fig.~4, similarly as in Fig. 2 for the case with two ISAC transmitters. Furthermore, it is observed that performance of SDR-based solution is closed to the CRLB-approximation-based solution  when $\Gamma$ is low, while the SDR-based solution outperforms the other two designs when $\Gamma$ becomes high. This shows that the SDR-based solution is most effective in power minimization while balancing the performance trade-off between sensing and communication.


\section{Conclusion}
This paper studied the coordinated power control in network ISAC, for the purpose of minimizing the total transmit power of multiple ISAC transmitters while ensuring the SINR constraints for communications and the CRLB constraint for estimation. We proposed two approaches, namely the SDR and the CRLB approximation, respectively, which  transform the original non-convex power minimization into convex forms, that can be solved efficiently. Numerical results show that the proposed SDR-based solution obtains the best performance as compared to the CRLB-approximation-based solution and a benchmark scheme with separate communication-sensing design. The investigation of coordinated  resource allocation for network ISAC under more complicated scenarios with, e.g., multiple antennas and wideband transmission, is interesting directions for future work.
\bibliographystyle{IEEEtran}


\begin{thebibliography}{1}

  \bibitem{OverviewCui}
  Y. Cui, F. Liu, X. Jing, and J. Mu, ``Integrating sensing and communications for ubiquitous IoT: Applications, trends and challenges,'' \emph{IEEE Network}, vol. 35, no. 5, pp.158-167, Sep./Oct. 2021.


  \bibitem{OverviewFan}
  F. Liu, Y. Cui, C. Masouros, J. Xu, T. X. Han, Y. C. Eldar, and S. Buzzi, ``Integrated sensing and communications: Towards dual-functional wireless networks for 6G and beyond,''   to appear in \emph{IEEE J. Sel. Areas Commun.}, [Online] Available: {\url{https://arxiv.org/pdf/2108.07165.pdf}}
  
  

  \bibitem{ZengJLC}
  Z. Xiao and Y. Zeng, ``An overview on integrated localization and communication towards 6G,'' \emph{Sci. China Inf. Sci.}, vol. 65, no. 131301,  pp. 1-46, Dec. 2022.

  \bibitem{Waveform}
  F. Liu, L. Zhou, C. Masouros, A. Li, W. Luo, and A. Petropulu, ``Toward dual-functional radar-communication systems: Optimal waveform design,'' \emph{IEEE Trans. Signal Process.}, vol. 66, no. 16, pp. 4264-4279, Aug. 2018.

  \bibitem{BeamOpt}
  H. Hua, J. Xu, and T. X. Han, ``Transmit beamforming optimization for integrated sensing and communication'',  in \emph{Proc. IEEE Global Commun. Conf. (GLOBECOM)}, Dec. 2021, pp. 1-6.

  \bibitem{Receiver}
  F. Liu, C. Masouros, A. P. Petropulu, H. Griffiths, and L. Hanzo, ``Joint radar and communication design: Applications, state-of-the-art, and the road ahead,'' \emph{IEEE Trans. Commun.}, vol. 68, no. 6, pp. 3834-3862, Jun. 2020.
  
  \bibitem{PerMobNet}
  A. Zhang, M. L. Rahman, X. Huang, Y. J. Guo, S. Chen, and R. W. Heath, ``Perceptive mobile networks: Cellular networks with radio vision via joint communication and radar sensing,'' \emph{IEEE Veh. Technol. Mag.}, vol. 16, no. 2, pp. 20-30, Jun. 2021.
  
  \bibitem{CramBoundGodrich}
  H. Godrich, A. M. Haimovich, and R. S. Blum, ``Target localization accuracy gain in MIMO radar based system,'' \emph{IEEE Trans. Inf. Theory}, vol. 56, no. 6, pp. 2783-2803, Jun. 2010.
  
 \bibitem{CoMP}
 D. Gesbert, S. Hanly, H. Huang, S. S. Shitz, O. Simeone, and W. Yu, ``Multi-cell MIMO cooperative networks: A new look at interference,'' \emph{IEEE J. Sel. Areas Commun.}, vol. 28, no. 9, pp. 1380-1408, Dec. 2010.
 
   \bibitem{PowerAlloGodrich}
 H. Godrich, A. P. Petropulu, and H. V. Poor, ``Power allocation strategies for target localization in distributed multiple-radar architectures,''  \emph{IEEE Trans. Signal Process.}, vol. 59, no. 7, pp. 3226-3240, Jul. 2011.
 


  \bibitem{PowerAllocaUAV}
  X. Wang, Z. Fei, J. A. Zhang, J. Huang, and J. Yuan, ``Constrained utility maximization in dual-functional radar-communication multi-UAV networks,'' \emph{IEEE Trans. Commun.}, vol. 69, no. 4, pp. 2660-2672, Apr. 2020.

  \bibitem{CramerRaoOpt}
  F. Liu, Y. Liu, A. Li, C. Masouros, and Y. C. Eldar, ``Cram\'{e}r-Rao bound optimization for joint radar-communication beamforming,'' \emph{IEEE Trans. Signal Process.}, vol. 70, pp. 240-253, Dec. 2021.

  \bibitem{SDR}
  Z. Luo, W. Ma, A. M. So, Y. Ye, and S. Zhang, ``Semidefinite relaxation of quadratic optimization problems,'' \emph{IEEE Signal Process. Mag.}, vol. 27, no. 3, pp. 20-34, May 2010.

  \bibitem{SDRboxCons}
  S. Burer and D. Vandenbussche, ``A finite branch-and-bound algorithm for nonconvex quadratic programming via semidefinite relaxations,'' \emph{Math. Program.}, vol. 113, no. 2, pp. 259-282, Dec. 2006.

  \bibitem{CVX}
  M. Grant and S. Boyd, ``CVX: Matlab software for disciplined convex programming (web page and software),'' \url{http://cvxr.com/cvx/}, Apr. 2010.

\end{thebibliography}
\bibliographystyle{IEEEtran}

\end{document}